\documentclass[conference]{IEEEtran}
\ifCLASSINFOpdf
  \usepackage[pdftex]{graphicx}
  \DeclareGraphicsExtensions{.pdf,.jpeg,.png}
\else
\fi

\newcommand {\apgt} {\ {\raise-.5ex\hbox{$\buildrel>\over\sim$}}\ }
\newcommand {\aplt} {\ {\raise-.5ex\hbox{$\buildrel<\over\sim$}}\ }

\newcommand{\mnras}{\mbox{MNRAS}}

\newcommand{\nat}{\mbox{Nature}}

\newcommand{\pasj}{\mbox{Publications of the Astronomical Society of Japan}}

\begin{document}
%
\title{From Thread to Transcontinental Computer: Disturbing Lessons in Distributed Supercomputing}

\author{\IEEEauthorblockN{Derek Groen*}
\IEEEauthorblockA{Centre for Computational Science and CoMPLEX\\
University College London\\
London, United Kingdom\\
Email: d.groen@ucl.ac.uk}
\and
\IEEEauthorblockN{Simon Portegies Zwart*}
\IEEEauthorblockA{Leiden Observatory\\
Leiden University\\
Leiden, the Netherlands\\
Email: spz@strw.leidenuniv.nl}}


%


\maketitle

\begin{abstract}

We describe the political and technical complications encountered during the
astronomical CosmoGrid project. CosmoGrid is a numerical study on the formation
of large scale structure in the universe. The simulations are challenging due
to the enormous dynamic range in spatial and temporal coordinates, as well as
the enormous computer resources required.  In CosmoGrid we dealt with the
computational requirements by connecting up to four supercomputers via an
optical network and make them operate as a single machine. This was
challenging, if only for the fact that the supercomputers of our choice are
separated by half the planet, as three of them are located scattered across
Europe and fourth one is in Tokyo.  The co-scheduling of multiple computers and
the 'gridification' of the code enabled us to achieve an efficiency of up to
$93\%$ for this distributed intercontinental supercomputer.  In this work, we
find that high-performance computing on a grid can be done much more
effectively if the sites involved are willing to be flexible about their user
policies, and that having facilities to provide such flexibility could be key
to strengthening the position of the HPC community in an increasingly
Cloud-dominated computing landscape. Given that smaller computer clusters owned
by research groups or university departments usually have flexible user
policies, we argue that it could be easier to instead realize distributed
supercomputing by combining tens, hundreds or even thousands of these
resources.


\end{abstract}


%
\IEEEpeerreviewmaketitle

\section{Introduction}

Computers have become an integral part of modern life, and are essential
for most academic research~\cite{Hettrick:2014}. Since the middle of
last century, researchers have invented new techniques to boost their
calculation rate, e.g. by engineering superior hardware, designing more
effective algorithms, and introducing increased parallelism. Due to a range of
physical limitations which constrain the performance of single processing
units~\cite{Markov:2014}, recent computer science research is frequently geared
towards enabling increased parallelism for existing applications.  

By definition, parallelism is obtained by concurrently using the calculation
power of multiple processing units. From small to large spatial scales, this is
respectively done by: facilitating concurrent operation of instruction threads
within a single core, of cores within a single processor, of processors within
a node, of nodes within a cluster or supercomputer, and of supercomputers
within a distributed supercomputing environment. The vision of aggregating
existing computers to form a global unified computing platform, and to focus
that power for a single purpose, has been very popular both in popular fiction
(e.g., the Borg collective mind in Star Trek or Big Brother in Orwell's 1984)
and in scientific research (e.g., Amazon EC2, projects such as
TeraGrid/XSEDE and EGI, and numerous distributed
computing projects~\cite{Gualandris:2007,Battestilli:2007,Balaji:2008,Groen:2011,Agullo:2011,Seinstra:2011,Belgacem:2013,Borgdorff:2014}).
Although many have tried, none have yet succeeded to link up more than a
handful of major computers in the world to solve a major high-performance
computing problem.

Very few research endeavors aim to do distributed computing at such a 
scale to obtain more performance. Although it requires a rather large labour investment 
across several time zones, accompanied with political complexities, it is technically
possible to combine supercomputers to form an intercontinental grid.
We consider that combining supercomputers in such a way is probably
worth the effort if many machines are involved, rather than a few.
Combining a small number of machines is hardly worth the effort of
doubling the performance of a single machine, but combining hundreds
or maybe even thousands of computers together could increase performance
by orders of magnitude~\cite{2008PCAA.book.....H}. 

Here we share our experiences, and lessons learned, in performing a large
cosmological simulation using an intercontinental infrastructure of multiple
supercomputers. Our work was part of the CosmoGrid
project~\cite{PortegiesZwart:2010,Groen:2011}, an effort that was eventually successful
but which suffered from a range of difficulties and set-backs. 
The issues we faced have impacted on our personal research ambitions, 
and have led to insights which could benefit researchers in any 
large-scale computing community.


We provide a short overview of the CosmoGrid project, and describe our initial
assumptions in Section~\ref{Sec:vision}.  We summarize the challenges we faced,
ascending the hierarchy from thread to transcontinental computer, in
Section~\ref{Sec:parallel} and we summarize how our insights affected our
ensuing research agenda in Section~\ref{Sec:after}. We discuss the long-term
implications of CosmoGrid in Section~\ref{Sec:future} and conclude the paper
with some reflections in Section~\ref{Sec:discuss}.

\section{The CosmoGrid project: vision and (implicit) assumptions}\label{Sec:vision}

The aim of CosmoGrid was to interconnect four supercomputers (one in Japan, and
three across Europe) using light paths and 10 Gigabit wide area networks, and to use them
concurrently to run a very large cosmological simulation. We performed the
project in two stages: first by running simulations across two supercomputers,
and then by extending our implementation to use four supercomputers concurrently.
The project started as a collaboration between researchers in the Netherlands,
Japan and the United States in October 2007, and received 
support from several major supercomputing centres (SARA in Amsterdam, EPCC in
Edinburgh, CSC in Espoo and NAOJ in Tokyo).  CosmoGrid mainly served a two-fold
purpose: to predict the statistical properties of small dark matter halos from an
astrophysics perspective, and to enable production simulations using an
intercontinental network of supercomputers from a computer science perspective.

\subsection{The software: GreeM and SUSHI}

For CosmoGrid, we required a code to model the formation of dark
matter structures (using $2048^3$ particles in total) over a period of over 13 
billion years. We adopted a hybrid Tree/Particle-Mesh (TreePM) N-body 
code named {\it GreeM}~\cite{Ishiyama:2009,Yoshikawa:2005}, which is highly scalable
and straightforward to install on supercomputers. GreeM uses a Barnes-Hut tree 
algorithm~\cite{Barnes:1986} to calculate force interactions between dark matter 
particles over short distances, and a particle-mesh algorithm to calculate force 
interactions over long distances~\cite{Hockney:1981}.
Later in the project, we realized that further code changes were required to
enable execution across supercomputers. As a result, we created a separate
version of GreeM solely for this purpose. This modified code is named \textit{SUSHI},
which stands for Simulating Universe Structure formation on Heterogeneous
Infrastructures~\cite{Groen:2011,Groen:2011-3site}.

\subsection{Assumptions}

Our case for a distributed computing approach was focused on a classic argument
used to justify parallelism: multiple resources can do more work than a single
one.  Even the world's largest supercomputer is about an order of magnitude
less powerful than the top 500 supercomputers in the world
combined~\cite{top500}. In terms of interconnectivity the case was also clear.
Our performance models predicted that a 1 Gbps wide area network would already
result in good simulation performance (we had 10 Gbps links at our disposal),
and that the round-trip time of about 0.27 s between the Netherlands and Japan
would only impose a limited overhead on a simulation that would require
approximately 100 s per integration time step.  World-leading performance of
our cosmological N-body integrator was essential to make our simulations
possible, and our Japanese colleagues optimized the code for single-machine
performance as part of the project. Snapshots/checkpoints would then be written
distributed across sites, and gathered at run-time.  At the start of CosmoGrid,
we anticipated to run across two supercomputers by the summer of 2008, and
across four supercomputers by the summer of 2009.

We assumed a number of political benefits: the simulation we proposed required
a large number of core hours and produce an exceptionally large amount of
data. These requirements would have been a very heavy burden for a single
machine, and by executing a distributed setup we could mitigate the
computational, storage and data I/O load imposed on individual machines. We
also were aware of the varying loads of machines at different times, and could
accommodate for that by rebalancing the core distribution whenever we would
restart the distributed simulation from a checkpoint.

Overall, we mainly expected technical problems,
particularly in establishing a parallelization platform which works across
supercomputers.  Installing homogeneous software across heterogeneous (and
frequently evolving) supercomputer platforms appeared difficult to accomplish,
particularly since we did not possess administrative rights on any of the
machines. In addition, the GreeM code had not
been tested in a distributed environment prior to the project.

\section{Distributed supercomputing in practice}\label{Sec:parallel}

Although we finalized the production simulations about a year later than
anticipated, CosmoGrid was successful in a number of fundamental areas. We
managed to successfully execute cosmological test simulations across up to four
supercomputers, and full-size production simulations across up to three
supercomputers~\cite{Groen:2011,Groen:2011-3site}. In addition, our
astrophysical results have led to new insights on the mass distribution of
satellite halos around Milky-Way sized galaxies~\cite{Ishiyama:2013}, on the
existence of small groups of galaxies in dark-matter deprived voids
\cite{2013MNRAS.435..222R}, the structure of voids \cite{2014arXiv1411.1276R}
and the evolution of barionic-dominated star clusters in a dark matter halo
\cite{2013MNRAS.436.3695R}. However, these results, though valuable in their own right, do not capture some
of the most important and disturbing lessons we have learned from CosmoGrid
about distributed supercomputing. Here we summarize our experiences on
engineering a code from the level of threads to that of a transcontinental
machine, establishing a linked infrastructure to deploy the code, reserving the
required resources to execute the code, and the software engineering and
sustainability aspects surrounding distributed supercomputing codes.

We are not aware of previous publications 
of practical experiences on the subject, and for that reason this paper may
help achieve a more successful outcome for existing research efforts in
distributed HPC. 

\subsection{Engineering a code from thread to transcontinental machine}

The GreeM code was greatly re-engineered during CosmoGrid. This was necessary to 
achieve a complete production simulation within the core hour allocations that we 
obtained from NCF and DEISA. Our infrastructure consisted of three Cray XT4 machines
with little Endian Intel chips and one IBM machine with Big Endian Power7 chips.
At the start of CosmoGrid, the Power7 architecture was not yet in place.  GreeM
had been optimized for the use of SSE and AVX instruction sets, executing 10
times faster when these instructions are supported. SSE was available in Intel
chips and AVX was expected to be available in Power7 chips. However, the
support for AVX in Power7 never materialized, forcing us to find alternative
optimization approaches. An initial 2-month effort by IBM engineers, leading to a 10\% performance
improvement, did not speed up the code sufficiently. We then resorted to 
manual optimization without the use specialized instruction sets, e.g., by 
reordering data structures and unrolling loops. This effort resulted in a 
$\sim 300\%$ performance increase, which was within a factor of 3 of our 
original target.
 
Much of the parellelization work on GreeM was highly successful, as evidenced
by the Gordon Bell Prize awarded in 2012 to Ishiyama et al.~\cite{Ishiyama:2012}. 
However, one unanticipated problem arose while scaling up the simulation 
to larger problem sizes. GreeM applies a Particle-Mesh (PM) algorithm to
resolve the interactions beyond a preconfigured cutoff limit. The implementation of 
is algorithm was initially serial, as the overhead was a negligible component ($<1$\%) of the total execution
time for smaller problems.  However, the overhead became much larger we scaled
up to mesh sizes beyond $256^3$ mesh cells, forcing us to move from a serial
implementation to a parallel implementation. 

\subsubsection{SUSHI and MPWide}

Initially we considered executing GreeM as-is, and using a middleware solution
to enable execution across supercomputers. In the years leading up to
CosmoGrid, a large number of libraries emerged for distributed HPC (e.g.,
MPICH-G2~\cite{Karonis:2003}, OpenMPI~\cite{openmpi-site,Coti:2009}, mpig~\cite{Manos:2008} and
PACX-MPI~\cite{Muller:2003}). Many of these were strongly recommended by
colleagues in the field, and provided MPI layers that allowed applications to
pass messages across different computational resources. Although these were
well-suited for distributed HPC across self-administered clusters, we quickly
found that a distributed supercomputing environment was substantially
different. 

First, supercomputers are both more expensive and less common than clusters, and
the centres managing them are reluctant to install libraries that require
administrative privileges, due to risks of security and ease of maintenance
(MPI distributions tend to require such privileges). Second, the networks
interconnecting the supercomputers are managed by separate organizational
entities, and the default configurations at each network endpoint are almost
always different and frequently conflicting. This is not the case in more
traditional (national) grid infrastructures, where uniform configurations can
be imposed by the overarching project team. The heterogeneity in network
configurations resulted in severe performance and reliability penalties when
using standard TCP-based middleware (such as MPI and {\tt scp}, unless we could
find some way to either (a) customize the network configuration for individual
paths or (b) adopt a different protocol (e.g., UDP) which ignores these preset
configurations. In either case, we realized that using standard TCP-based MPI 
libraries for the communication between supercomputers was no longer a viable 
option.

Using any other library had the inevitable consequence of modifying the main
code, and eventually we chose to customize GreeM (the
customized version is named SUSHI~\cite{Groen:2011-3site}) and
establish a seperate communication library
(MPWide~\cite{Groen:2010,Groen:2013}) for distributed supercomputing.

\subsection{Establishing a distributed supercomputing infrastructure}

Having a code to run, and computer time to run it on is insufficient to do
distributed concurrent supercomputing. The amount of data to be transported is
$\sim 10\,Gb$ per integration time step and should not become the limiting
factor in measuring performance. Each integration step would take about 100\,s.
When allowing a 10\% overhead we would have to require a network speed of
$\apgt 1$\,Gb/s. Our collaboration with Cees de Laat (University of Amsterdam)
and Kei Hiraki (Tokyo University) enables us to have two network and data
transport specialists at each side of the light path.

\begin{figure}[!t]
\centering
\includegraphics[width=3in]{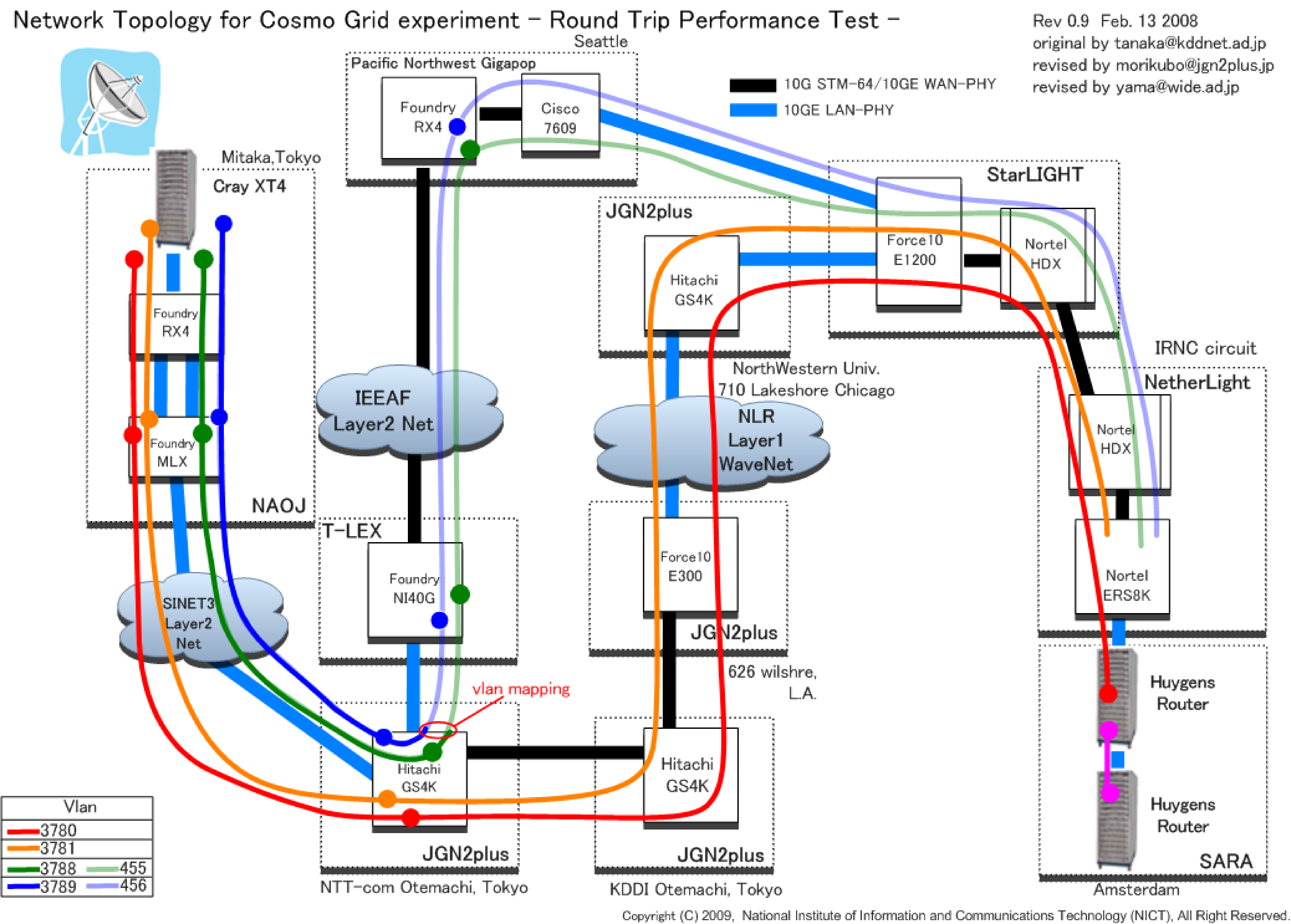}

\caption{Network topology map of the Amsterdam-Tokyo light path, shown with
reduced complexity, as it was envisioned in 2009. The revision number
(v0.9) is given in the top corner (together with an incorrect date), 
correctly implying that throughout the project we updated this 
topology map at least eight times. Reproduced from the preprint version 
of Portegies Zwart et al.~\cite{PortegiesZwart:2010}.} 

\label{Fig:CGNetworkTopology}
\end{figure}

At the time, Russia had planned to make their military 10\,Gbps dark
fiber available for scientific experiments, but due to their enhanced
military use we were unable to secure access to this cable. The
eventual route of the optical cable is presented in
Fig.\,\ref{Fig:CGNetworkTopology}.  We had a backup network between the
NTT-com router at JGN2plus and the StarLIGHT interconnect to guarantee
that our data stream remained stable throughout our calculations.

One of the interesting final quirks in our network topology was the
absence of an optical network interface in the Edinburgh machine
(which was installed later), and the fact that the optical cable at
the Japanese side reached the computer science building on the Mitaka
campus in Tokyo next to where the supercomputer at NAOJ was located. A
person had to go physically to dig a hole and install a connecting
cable between the two buildings.

From a software perspective, we present our design considerations on MPWide
fully in Groen et al.~\cite{Groen:2013}. Here we will summarize the main
experiences and lessons that we learned from CosmoGrid, as well as our 
experiences in readying the network in terms of software configuration.
We initially attempted to homogenize the network
configuration settings between the different supercomputers. This effort failed, as it
was complicated by the presence of over a dozen stakeholder organizations, and
further undermined by the lack of diagnostic information available to us. For
example, it was not always possible for us to pinpoint specific configuration
errors to individual routers, and to their respective owners. We also assessed
the performance of UDP-based solutions such as Quanta~\cite{He:2003} and
UDT~\cite{Gu:2007}, which operate outside of the TCP-specific configuration
settings of network devices. However, we were not able to universally adopt
such a solution, as some types of routers filter or restrict 
UDP traffic.

\begin{figure}[!t]
\centering
\includegraphics[width=3in]{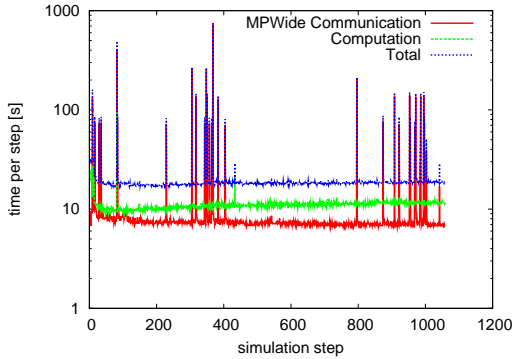}

\caption{Example of glitches observed in the communication path. This is a
performance measurements of a cosmological simulation using $256^3$ particles,
run in parallel across the supercomputers in Amsterdam and Tokyo (32 cores per
site). We present the total time spent (blue dotted line), time spent on
calculation (green dashed line), and the time spent on communication with
MPWide (red solid line). Stalls in the communication due to dropped packets
resulted in visible peaks in the communication performance measurements.
Reproduced from Groen et al.~\cite{Groen:2010}. Note: The communication time
was relatively high ($\sim 35\%$ of total runtime)) in this test run due to the
small problem size, in~\cite{Groen:2010} we also present results from a test
run with $2048^3$ particles, which had a communication overhead of $\sim 13\%$
of total runtime.} 

\label{Fig:CGglitch}
\end{figure}

We eventually converged on basing MPWide on multi-stream
TCP~\cite{Hacker:1993}, and combine this with mechanisms to customize TCP
settings on each of the communication nodes~\cite{Groen:2013}. Our initial
tests were marred with network glitches, particularly on the path between
Amsterdam and Tokyo (see Fig.~\ref{Fig:CGglitch} for an example where packets
were periodically stalled). However, later runs resulted in more stable
performance once we used a different path and adjusted the MPWide configuration
to use very small {\tt TCP} buffer sizes per stream~\cite{Groen:2011}.

\subsection{Reserving supercomputers}

The ability to have concurrent access to multiple supercomputers is an essential
requirement for distributed supercomputing. Within CosmoGrid, we initially 
agreed that the four institutions involved would provide so-called "phone-based" 
advance reservation for the purpose of this project. However, due to delays in
the commisioning of the light path, and due to political resistance regarding advance
reservation within some of the supercomputing centres (in part caused due to increasing
demand of the machines), it was no longer possible to use this means of 
advance reservation on all sites. Eventually, we ended up with a different ``reservation''
mechanism for each of the supercomputers.

The original approach of calling up was still supported for the
Huygens machine in the Netherlands. For the Louhi machine in Finland, calling
up was also possible, but the reservation was established through an
exclusively dedicated queue, as no direct reservation system was in place. This
approach had the side effect of locking other users out, and therefore we only
opted to use it as a last resort. For the HECToR machine in the UK, it was
possible to request higher priority, but not to actually reserve resources in
advance. Support for advance reservation was provided there shortly after CosmoGrid concluded,
but at the time the best method to ``reserve'' the machine was to submit a
very large high priority job right before the machine maintenance time slot.
We then need to align all other reservations to the end of that maintenance time slot,
presumed usually to be 6 hours after the start of the maintenance. For the CRAY
machine in Japan, reservation was no longer possible due to the high work load.
However, some mechanisms of augmented priority could be established indirectly,
e.g. by chaining jobs, which allowed for a job to be kept running at all times.

The combination of these strategies made it impossible to perform a large production run using all
four sites. We did perform smaller tests using the full infrastructure (using regular scheduling 
queues and hoping for the best), but we were only able to do the largest runs using either the 
three European machines, or Huygens combined with the CRAY in Tokyo.

\subsection{Software engineering and sustainability}

The task of engineering a code for distributed supercomputing is accompanied
with a number of unusual challenges, particularly in the areas of software
development and testing~\cite{Derek-SSI:2013}. At the time, we had to ensure
that GreeM and SUSHI remained fully compatible with all four supercomputer
platforms. With no infrastructure in place to do continuous integration testing
on supercomputer nodes (even today this is still a rarity), we performed this
testing periodically by hand. In general, testing on a single site is
straightforward, but testing across sites less so. 

We were able to arrange proof-of-concept tests across 4 sites using very small
jobs (e.g., 16 cores per site) by putting these jobs with long runtimes in the
queue on each machine and waiting with starting the run until the jobs are
running simultaneously.  For slightly larger jobs (e.g., 64 cores per site)
this became difficult during peak usage hours as the queuing times of
each job became longer and less predictable. However, we have been able to perform a
number of tests during more quiet periods of the week (e.g., at 2 AM), without
using advance reservation. For yet larger test runs we were required to use
advance reservation, reduce the number of supercomputers involved, or both.

Software testing is instrumental to overall development, particularly when
developing software in a challenging environment like an intercontental network
of supercomputers. Here, the lack of facilities for advance reservation and
continuous integration made testing large systems prohibitively difficult, and
had an adverse effect on the development progress of SUSHI. We eventually
managed to get an efficient production calculation running for 12 hours across
three sites and with the full system size of $2048^3$
particles~\cite{Groen:2011-3site}, but with better facilities for testing
across sites we could well have tackled larger problems using higher core counts.

More emphasis and investment in testing facilities at the supercomputer centres
would have boosted the CosmoGrid project, and such support would arguably
be of great benefit to increase the user uptake of supercomputers in general.

\begin{figure}[!t]
\centering
\includegraphics[width=3in]{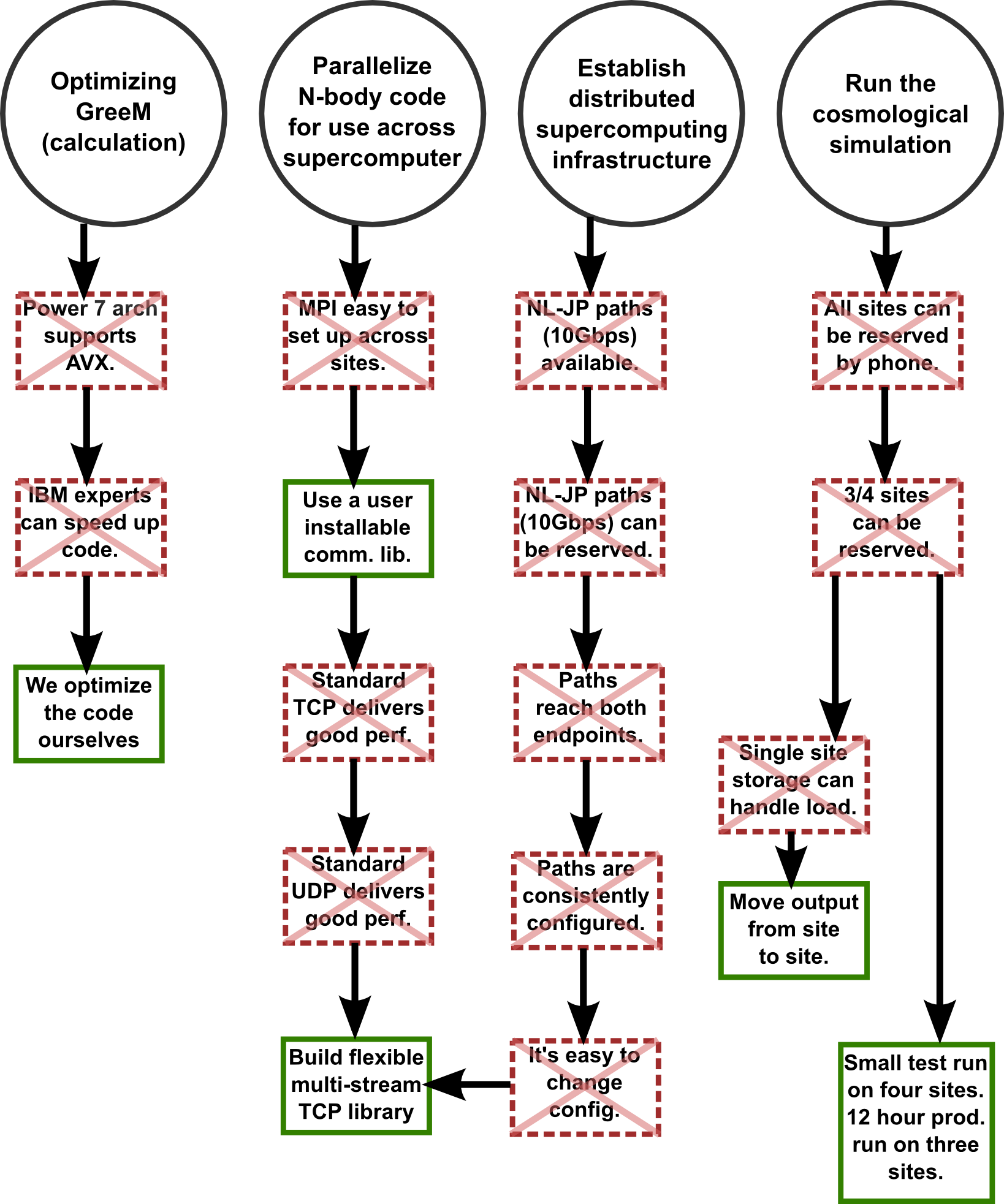}

\caption{Overview of situations where the CosmoGrid project deviated from the
original planning. We summarize the four main computational activities in the
circles at the top, and provide examples of assumptions, which were shown to be
incorrect during the project, using red boxes with transparent crosses. We
present examples of effective workarounds using green boxes. We provide further
details on each situation in Sec.~\ref{Sec:parallel}.}

\label{Fig:CGFlowchart}
\end{figure}

\section{Research directions after CosmoGrid}\label{Sec:after}

Our experience with CosmoGrid changed how we approached our computational 
research in two important ways. First, due to the political hardships and
the lack of facilities for advance reservation and testing, we changed our 
emphasis from distributed concurrent supercomputing towards improving 
code performance on single sites~\cite{Ishiyama:2012,Bedorf:2014:PGT:2683593.2683600}.

Second, our expertise enabled us the enter the relatively young field of
distributed multiscale computing with a head start. Multiscale computing
involves the combination of multiple solvers to solve a problem that
encompasses several different length and time scales. Each of these solvers may
have different resource requirements, and as a result such simulations are
frequently well-suited to be run across multiple resources. Drost et al.
applied some of our expertise, combining it with their experience using IBIS,
to enable the AMUSE environment to run different coupled solvers on different
resources concurrently~\cite{Drost:2012}.

Our experiences with CosmoGrid were an important argument towards redeveloping
the MUSCLE coupling environment for the MAPPER project. In this EU-funded 
consortium, Borgdorff et al. developed a successor (MUSCLE 2), which is optimized 
for easier installation on supercomputers, and automates the startup of 
solvers that run concurrently on different sites (a task that was done manually
in CosmoGrid). In addition, we integrated MPWide in MUSCLE2~\cite{Borgdorff:2014-2} and 
used MPWide directly to enable concurrently running coupled simulations across 
sites~\cite{Groen:2013-2}.

\section{Future prospects of distributed supercomputing}\label{Sec:future}

Distributed high-performance supercomputing is not for the faint at
heart. It requires excellent programming and pluralization skills,
stamina, determination, politics and hard labour.

One can wonder if it is worth the effort, but the answer depends on the
available resources and the success of proposal writing.  About 20--30\% of the
proposals submitted to INCITE (http://www.doeleadershipcomputing.org/faqs/) 
or PRACE (http://www.prace-ri.eu/prace-kpi/) are successful, but these
success rates are generally lower for the very largest infrastructures 
(e.g., ORNL Titan). In addition, some of the largest supercomputers (such as Tianhe-2
and the K Computer) provide access only to closely connected research groups or
to projects that have a native Principal Investigator. Acquiring compute time
on these largest architectures can be sufficiently challenging that running
your calculations on a number of less powerful but earlier accessible machines
may be easier to accomplish.

Accessing several of such machines through one project is even harder,
and probably not very realistic. Similarly, for the different
architectures, it would be very curious to develop a code that works
optimal on K computer and ORNL Titan concurrently. Achieving 25 PFlops
on Titan alone is already a major undertaking
\cite{Bedorf:2014:PGT:2683593.2683600}, and combining such an
optimized code with a Tofu-type network architecture (which is 
present on the K computer) would make optimization a challenge 
of a different order. 

We therefore do not think that distributed architectures will be used
to beef-up the world's fastest computers, nor to connect a number of
top-10 to top-100 supercomputers to out-compute the number 1. The type
of distributed HPC as discussed in this article is probably best
applied to large industrial or small academic computer clusters. These
$\aplt 2$\,PetaFlop architectures are found in many academic settings or
small countries, and are relatively easily accessible, by
peer review proposals or via academic license agreements.  In this context,
we think it is more feasible to connect 10 to 100 of such machines to
outperform a top 1 to 10 computer.

\section{Conclusions}\label{Sec:discuss}

We have presented our experiences from the CosmoGrid project in
high-performance distributed supercomputing. Plainly put, distributed
high-performance supercomputing is a tough undertaking, and most of our 
initial assumptions were proven wrong. Much of the
hardware and software infrastructure was constructed with very specific
use-cases in mind, and was simply not fit for purpose to do distributed
high-performance supercomputing. A major reason why we have been able to
establish distributed simulations at all is due to the tremendous effort of all
the people involved, from research groups, networking departments and
supercomputer centres. It was due to their efforts to navigate the project
around the numerous technical and political obstacles that distributed 
supercomputing became even possible.

CosmoGrid was unsuccessful in establishing high-performance distributed
supercomputing as the future paradigm for using very large supercomputers.
However, the project did provide a substantial knowledge boost to our
subsequent research efforts, which is reflected by the success of projects such
as MAPPER~\cite{Borgdorff:2014,Groen:2013-2} and
AMUSE~\cite{Drost:2012,PortegiesZwart:2013}. The somewhat different approach
taken in these projects (aiming for more local resource infrastructures, and
with a focus on coupling different solvers instead of parallelizing a single
one) resulted in tens of publications which relied on distributed
(super-)computing. 

The HPC community has recently received criticism for its conservative
approaches and resistance to change (e.g.,~\cite{Dursi:2015,Squyres:2015}).
Through CosmoGrid, it became obvious to us that resource providers can be
subject to tricky dilemmas, where the benefits of supporting one research
project need to be weighed against the possible reduced service (or support)
incurred by other users. In light of that, we do understand the conservative
approaches followed in HPC to some extent. In CosmoGrid, we tried to work
around that by ensuring that our software was installable without any
administrative privileges, and we recommend that new researchers who wish to do
distributed supercomputing do so as well (or, perhaps, adopt very robust,
flexible and well-performing tools for virtualization). 
In addition, we believe that CosmoGrid would have been greatly helped if 
innovations such as automated advance reservation
systems for resources and network links, facilities for systematic software
testing and continuous integration, and streamlined procedures for obtaining
access to multiple sites had been in place. Even today, such facilities make 
HPC infrastructures more convenient for new types of users and applications,
and strengthen the position of the HPC community in an increasingly 
Cloud-dominated computing landscape.

\section*{Acknowledgment}

Both SPZ and DG contributed equally to this work. We are grateful to Tomoaki Ishiyama, Keigo
Nitadori, Jun Makino, Steven Rieder, Stefan Harfst, Cees de Laat, Paola Grosso,
Steve MacMillan, Mary Inaba, Hans Blom, Jeroen B\'edorf, Juha Fagerholm, Tomoaki Ishiyama, Esko
Ker\"anen, Walter Lioen, Jun Makino, Petri Nikunen, Gavin Pringle and Joni
Virtanen for their contributions to this work.
This research is supported by the Netherlands organization for Scientific
research (NWO) grant \#639.073.803, \#643.200.503 and \#643.000.803 and the
Stichting Nationale Computerfaciliteiten (project \#SH-095-08). We thank
the DEISA Consortium (EU FP6 project RI-031513 and FP7 project RI-222919)
for support within the DEISA Extreme Computing Initiative (GBBP project).
This paper has been made possible with funding from the UK Engineering
and Physical Sciences Research Council under grant number EP/I017909/1
(http://www.science.net).



%

\end{document}